\documentclass[12pt]{article}
\usepackage{a4wide}
\usepackage{amsmath}
\usepackage{amssymb}
\usepackage{amsopn}

\begin{document}
{\renewcommand{\thefootnote}{\fnsymbol{footnote}}
\hfill  AEI--2004--048\\ 
\medskip
\hfill gr--qc/0407017\\
\medskip
\begin{center}
{\LARGE  Spherically Symmetric Quantum Geometry:\\ States and Basic Operators}\\
\vspace{1.5em}
Martin Bojowald\footnote{e-mail address: {\tt mabo@aei.mpg.de}}
\\
\vspace{0.5em}
Max-Planck-Institut f\"ur Gravitationsphysik, Albert-Einstein-Institut,\\
Am M\"uhlenberg 1, D-14476 Golm, Germany
\vspace{1.5em}
\end{center}
}

\setcounter{footnote}{0}

\newtheorem{lemma}{Lemma}

\newcommand{\proofend}{\raisebox{1.3mm}{\fbox{\begin{minipage}[b][0cm][b]{0cm}
\end{minipage}}}}
\newenvironment{proof}{\noindent{\it Proof:} }{\mbox{}\hfill \proofend\\\mbox{}}

\newcommand{\case}[2]{{\textstyle \frac{#1}{#2}}}
\newcommand{\lP}{\ell_{\mathrm P}}

\newcommand{\md}{{\mathrm{d}}}
\newcommand{\Kern}{\mathop{\mathrm{ker}}}
\newcommand{\tr}{\mathop{\mathrm{tr}}}
\newcommand{\sgn}{\mathop{\mathrm{sgn}}}

\newcommand*{\R}{{\mathbb R}}
\newcommand*{\N}{{\mathbb N}}
\newcommand*{\Z}{{\mathbb Z}}
\newcommand*{\Q}{{\mathbb Q}}
\newcommand*{\C}{{\mathbb C}}
\newcommand{\abs}[1]{\lvert#1\rvert}

\newcommand{\Lam}{\Lambda}
\newcommand{\vt}{\vartheta}
\newcommand{\vp}{\varphi}
\newcommand{\Ab}{\ensuremath{\bar{\cal A}_{\rm sph.symm.}}}

\begin{abstract}
 The kinematical setting of spherically symmetric quantum geometry,
 derived from the full theory of loop quantum gravity, is
 developed. This extends previous studies of homogeneous models to
 inhomogeneous ones where interesting field theory aspects arise. A
 comparison between a reduced quantization and a derivation of the
 model from the full theory is presented in detail, with an emphasis
 on the resulting quantum representation. Similar concepts for
 Einstein--Rosen waves are discussed briefly.
\end{abstract}

\section{Introduction}

Since general relativity predicts singularities generically, and in
particular in physically interesting situations such as cosmology and
black holes, it cannot be complete as a physical theory. The situation
improves when one quantizes general relativity in a background
independent manner, following loop quantum gravity \cite{Rev}. The
dynamics of the full theory is not yet settled and is rather
complicated, as expected for a full quantum theory of gravity. Even
classically one usually introduces symmetries for physical
applications, which can also be done in loop quantum gravity
directly. This in fact lead to the conclusion that isotropic models in
loop quantum cosmology \cite{IsoCosmo} are non-singular
\cite{Sing} while at the same time they show the usual classical
behavior at large scales \cite{SemiClass}.

One has to keep in mind, though, that symmetric models in a quantum
theory play a role different from symmetric classical solutions. While
the latter are exact solutions of the full theory, the former are
obtained from the full theory by completely ignoring many degrees of
freedom which violates their uncertainty relations. One thus should
weaken the symmetry by looking at less symmetric models, and check if
results obtained are robust. For the isotropic results, this has been
shown to be the case in a first step, reducing the isotropy by using
anisotropic but still homogeneous models \cite{HomCosmo,Spin}. This
did not only show that the same mechanism for singularity freedom
applied, and this in a more non-trivial way, but also lead to new
applications \cite{NonChaos}. The latter allow tentative conclusions
even for general, inhomogeneous singularities \cite{ChaosLQC}.

Nevertheless, one should go ahead and reduce the symmetry further. The
next step must deal with inhomogeneous models, which for simplicity
can first be taken to be $1+1$ dimensional. This would also allow new
physical applications concerning, e.g., spherically symmetric black
holes and cylindrical gravitational waves. Furthermore, they allow
additional tests of issues in the full theory which trivialize in
homogeneous models, including field theory aspects, the constraint
algebra and the role of anomalies, and specific constructions of
semiclassical states using graphs. For $1+1$ dimensional models
several alternative background independent quantization schemes have
been applied, which can then be compared with loop results. The
spherically symmetric model has been dealt with in the Dirac program
\cite{SphKl1} as well as in a reduced phase space quantization
\cite{Kuchar,SphKl2}. Einstein--Rosen waves can be mapped to a free
field on flat space-time allowing standard Fock quantization
techniques \cite{K:EinsteinRosen}, and there are several other
interesting models with a two-dimensional Abelian symmetry group which
have been quantized and studied extensively
\cite{AP:EinsteinRosen,LargeEff,MicroCyl,Planar,Gowdy,CylWaves,AsympSafe}. A
wide class of models, which have finitely many physical degrees of
freedom and also include the spherically symmetric model, is given by
dilaton gravity in two dimensions \cite{DilatonRev} or, more
generally, Poisson Sigma Models \cite{PSM}. These models have
been quantized exactly in a background independent way with reduced
phase space, Dirac or path integral methods.

The reason for the simplification in homogeneous models, which lead to
explicit cosmological applications, is not just the finite number of
degrees of freedom, but also a simplification of the volume operator
(which at first sight is not always explicit \cite{cosmoII}).  In
isotropic as well as diagonal homogeneous models the volume spectrum
can be computed explicitly, which is not possible in the full
theory. Since the volume operator plays a major role in defining the
dynamics \cite{QSDI}, also the evolution equation can be obtained and
analyzed in an explicit form. One can see that this is a consequence
of either a non-trivial isotropy subgroup of the symmetry group, or of
a diagonalization condition. Similar simplifications can be expected
more generally, in particular in those inhomogeneous models which have a
non-trivial isotropy group (spherical symmetry) or a diagonalization
condition on the basic variables (polarized waves).

Nevertheless, the explicit reduction of spherically symmetric models
done later in this paper, and also that of polarized cylindrical
waves, shows that suitable canonical variables display a feature
different from both the full theory and from homogeneous models: flux
variables (canonical momenta of the connection) are not identical to
the densitized triad which contains all information about spatial
geometry. (A similar feature, though in a different manner, happens in
the full theory when a scalar is coupled non-minimally
\cite{NonminScalar}.) Instead, the triad is a rather complicated
function of the basic variables and in particular depends also on the
connection. This seems to lead to an unexpected complication for the
volume operator, and shows that $1+1$ models are more complicated than
homogeneous ones not just for the obvious reason of having infinitely
many kinematical degrees of freedom, but also due to their canonical
structure. The complicated expression for the triad could, a priori,
even lead to a continuous volume spectrum, which would be difficult to
reconcile with the full theory and homogeneous models.  We will deal
with the volume operator elsewhere \cite{SphSymmVol}, but already in
this paper, where we introduce the kinematical setup and discuss
states and basic operators for connections and their momenta, we can
see that this issue can have an influence on semiclassical properties.

We will start by recalling the definition of symmetric states in a
quantum theory of connections, and then reduce the full phase space to
that of the spherically symmetric sector. We introduce the states of
the model by two procedures, first by loop quantizing the classically
reduced phase space and then by reducing states of the full theory to
be spherically symmetric. Both procedures lead to the same result,
which is a ``mixed'' quantization based on generalized connections, as
in the full theory \cite{ALMMT}, as well as elements of the Bohr
compactification of the real line, which is characteristic for
homogeneous models \cite{Bohr}. Quantum numbers of the reduced
quantization match with the spin labels obtained by restricting full
states, and gauge invariant reduced states satisfy the reduced Gauss
constraint. Basic operators on those states are given by holonomies
and fluxes, which suggest conditions for the semiclassical
regime. Finally, we will briefly discuss the model of Einstein--Rosen
waves.

\section{Symmetric states}

Let $\Sigma$ be a manifold carrying an action of a symmetry group $S$
such that there is a dense subset of $\Sigma$ where the group action
has an isotropy subgroup isomorphic to $F<S$. In this case $\Sigma$,
except for isolated points (symmetry axes or centers), can be
decomposed as $\Sigma\cong B\times S/F$ with the reduced manifold
$B=\Sigma/S$. On the symmetry orbits $S/F$ there is a natural
invariant metric which follows from the transitive group action, as
well as preferred coordinates.  On $B$, on the other hand, there is no
natural metric and no preferred coordinates.

A given symmetry group $S$ acting on a manifold $\Sigma$ defines a
class of inequivalent principal fiber bundles $P(\Sigma,G)$, for a
given group $G$, which carry a lift of the action of $S$ from $\Sigma$
to $P$ \cite{KobNom,Brodbeck}. For each such symmetric bundle there is
a set of invariant connections having a ${\cal L}G$-valued 1-form on
$P$ satisfying $s^*\omega=\omega$ for each $s\in S$, giving rise to
different embeddings $r_k\colon {\cal A}_{\rm inv}^{(k)}\to {\cal A}$
in the full space of connections. Here, $k$ is a label (topological
charge) characterizing the type of symmetric bundle used. In a
gravitational situation, where there is an additional condition that
spaces of connections must allow non-degenerate dual vector fields,
the would-be non-degenerate triads, usually only one value for the
label $k$ can be used such that we will suppress it later on.

An invariant connection then has the general form $A=A_B+A_{S/F}$
where $A_B$ is a reduced connection over $B$, in general with a
reduced gauge group, and $A_{S/F}$ contains additional fields in an
associate bundle transforming as scalars taking values in a certain
representation of the reduced structure group. The different forms
$r_k$ of embedding invariant connections into the full space of
connections are classified by homomorphisms $\lambda_k\colon F\to G$
up to conjugacy in $G$. This map also determines the reduced structure
group for the connections $A_B$ as the centralizer $Z_G(\lambda_k(F))$
in $G$. The additional fields in $A_{S/F}$ are the components of a
linear map $\phi\colon{\cal L}F_{\perp}\to{\cal L}G$ where the space
$S/F$ is assumed to be reductive, i.e., there is a decomposition
${\cal L}S={\cal L}F\oplus {\cal L}F_{\perp}$ such that ${\cal
L}F_{\perp}$ is fixed by the adjoint action of $F$. There are
additional linear conditions $\phi$ has to satisfy when it comes from
a full connection, namely
\begin{equation} \label{scalrel}
 \phi({\rm Ad}_fX)={\rm Ad}_{\lambda_k(f)}\phi(X)
\end{equation}
for all $X\in {\cal L}F_{\perp}$ and $f\in F$.

\subsection{Reduced loop quantization}

Loop quantum gravity provides techniques to quantize theories of
connections, possibly coupled to other fields, in a background
independent manner. Following this procedure, the component $A_B$,
which plays the role of the connection of the reduced theory, will be
quantized by using its holonomies along curves in $B$ as basic
variables \cite{ALMMT}. This leads to the space of generalized reduced
connections, $\bar{\cal A}_B$. Scalars like those in $A_{S/F}$ can be
quantized according to \cite{FermionHiggs,ScalarBohr} with the result
that the classical real values of the field are replaced by values in
the Bohr compactification of the real line. In this way, the space
$\bar{\cal A}_{B\times S/F}$ of generalized connections and scalar
fields becomes a compact group which carries a Haar measure
$\mu_0$. The Hilbert space $L^2(\bar{\cal A}_{B\times S/F},\mu_0)$ is
then obtained by completing the space of continuous functions on this
group with respect to the Haar measure.

Holonomies of the connection and analogous expressions for the scalar
act as multiplication operators, while the momenta of the connection
components, which can be written as fluxes, act as derivative
operators. Both sets of basic operators are subsequently used to
quantize more complicated, composite expressions.

\subsection{Symmetric states from the full theory}

Using the connection representation of states on the space of
generalized connections, symmetric states can be defined in the full
theory as distributional states supported only on invariant
connections \cite{SymmRed,PhD} for a given symmetry. It is clear that such
a state can also be represented as a function $\psi$ on the space of
reduced connections as before, but in addition it acquires an
interpretation as a distribution in the full theory, i.e.\ as a linear
functional on the space of cylindrical states depending only on
finitely many holonomies and scalar values: for any cylindrical
function $f$ on the full space of connections,
\begin{equation} \label{dist}
 \Psi[f]:=\int_{\bar{{\cal A}}^{(k)}_{B\times S/F}}\md\mu_0(A)
\overline{\psi(A)} \cdot r_k^*f(A)
\end{equation}
defines a distribution in the full theory. Symmetric states thus form
a subspace of the full distributional space which, using the measure on
$\bar{{\cal A}}^{(k)}_{B\times S/F}$ can be equipped with an inner
product.

Operators $\hat{O}$ of the full theory act on distributions $\Psi$ via the
dual action which defines $\hat{O}\Psi$ by
\[
 \hat{O}\Psi[f]=\Psi[\hat{O}^{\dagger}f] \quad\mbox{for all } f\in{\rm
Cyl}\,.
\]
In general, however, $\hat{O}\Psi$ will not be a symmetric state even
if $\Psi$ is. The reason is that on states in the connection
representation only the condition of having invariant $A$ has been
incorporated, but not the condition for invariant momenta $E$. Then,
even classically the flow generated by a phase space function would in
general not be tangential to the subspace given by invariant
connections and arbitrary triads.

For general operators it is therefore necessary to implement the
condition for invariant triads, which must be done by modifying the
operators suitably. This is a complicated procedure which has not been
developed in detail yet. Fortunately, one can use particular
operators in the full theory whose dual action leaves the space of
symmetric states invariant such that one can directly use them in the
reduced model. Classical analogs of those operators generate a flow
which is tangential to the subspace of invariant connections in phase
space even if triads can be arbitrary. It is easy to see that such
functions have to be linear in the triads (which is, however, not a
sufficient condition). In fact, the reduced basic variables,
holonomies and fluxes, are linear in the triads, and can be written
such that they generate a flow parallel to invariant
connections. Moreover, for the basic quantities, the classical Poisson
*-algebra is represented faithfully on the Hilbert space such that the
classical flow on phase space corresponds to a unitary transformation
in the quantum theory. Thus, quantizations of basic variables will map
symmetric states to symmetric states and can be used directly to
derive the reduced operators. States as well as basic operators of a
model are then defined directly in the full theory, and more
complicated operators can be constructed from the basic ones following
the lines of the full theory. An advantage of relating the model to
the full theory in this way is that there is a unique (under weak
conditions) diffeomorphism invariant representation of the full
holonomy/flux algebra \cite{LOST}
while within models one usually has several options.

The reduced theory is usually not a pure gauge theory even in the
absence of matter since some components $\phi$ of the full connection
play the role of scalar fields in the reduced model. Another
difference to the full theory is that often the model has a reduced
gauge group. States of the full theory and the model are then based on
different groups. Nevertheless, representations of the reduced
structure group automatically occur when the reduction from full
states is done. For an explicit representation of states and operators
an Abelian gauge group is most helpful since all irreducible
representations are then one-dimensional and there are no complicated
coupling coefficients between different representations. In fact, a
non-trivial isotropy subgroup of the symmetry group often, as in the
spherically symmetric case, implies an Abelian gauge group. Similarly,
diagonalization conditions imposed on connections and triads can lead
to Abelian gauge groups. On the other hand, a non-trivial isotropy
group or additional diagonalization conditions lead to additional
complications since the relations (\ref{scalrel}) have to be taken
care of. Moreover, even though the reduced connection may be Abelian,
its holonomies do in general not commute with expressions (point
holonomies) representing scalars since this would be incompatible with
a non-degenerate triad. Simplifications of an Abelian theory, such as
spin network states with an Abelian group, then are not always
obvious.

In many cases, the combined system of the reduced Abelian connection
plus scalar fields $\phi$ can be simplified taking into account the
special form of a given class of invariant connections.  A model can
be formulated as essentially Abelian if connection components along
independent directions along $B$ and in the orbits are perpendicular
in the Lie algebra. For instance, in the $1+1$ dimensional case, a
connection in general has the form
\begin{equation} \label{Agen}
 A=A_x(x) \Lambda_x(x) \md x+ A_y(x) \Lambda_y(x) \md y+ A_z(x)
\Lambda_z(x) \md z+ \mbox{field independent terms}
\end{equation}
where $x$ is the inhomogeneous coordinate on $B$ and $\Lambda_I(x)\in
{\cal L}G$. (Depending on the symmetry, there can be additional terms
not depending on fields $A_I$, as happens in the spherically symmetric
case discussed later.) The fields $A_y$ and $A_z$ together with
components of $\Lambda_y$ and $\Lambda_z$ comprise the field $\phi$
determining $A_{S/F}$ and are thus subject to
(\ref{scalrel}). Simplifications occur if we have
$\tr(\Lambda_x\Lambda_y)= \tr(\Lambda_x\Lambda_z)=
\tr(\Lambda_y\Lambda_z)=0$, as happens in the spherically symmetric
case or for cylindrical gravitational waves with a polarization
condition. Then, holonomies of invariant connections take a
corresponding form with perpendicular internal directions, and thus
obey special relations that would not hold true for holonomies of an
arbitrary connection. The most important relation which will be used
later is

\begin{lemma} \label{lemma}
 Let $g:=\exp(A)$ and $h:=\exp(B)$ with $A,B\in{\rm su}(2)$ such that
$\tr(AB)=0$. Then
\begin{equation} \label{holrel}
 gh=hg+h^{-1}g+hg^{-1}-\tr(hg)\,.
\end{equation}
\end{lemma}

\begin{proof} 
Since the equation (\ref{holrel}) is invariant under conjugation of
both $g$ and $h$ with the same SU(2)-element, we can first rotate $A$
to equal $a\tau_1$ for some $a\in\R$. Then, $\tr(AB)=0$ implies that
$B= b_2\tau_2+b_3\tau_3$ which can be rotated to $B=b\tau_2$ while
keeping $A$ fixed.

The proof proceeds by directly computing all products involved, using
$\exp(a\tau_i) =\cos(a/2)+2\tau_i\sin(a/2)$.
\end{proof}

Thus, even though reduced holonomies do not all lie in an Abelian
subgroup, they are almost commuting in the sense that products of two
holonomies can always be expanded into terms where the order is
reversed. It turns out that this is sufficient for a simplification of
the representation of states and basic operators, and in turn of other
ones like the volume operator. This has been exploited in homogeneous
models, where the special form of connections was a consequence of the
non-trivial isotropy \cite{IsoCosmo} or a diagonalization condition
\cite{HomCosmo}. Similarly in $1+1$ dimensional models, a non-trivial
isotropy group or a diagonalization condition can lead to connection
components which are perpendicular for independent directions. As we
will discuss in what follows, this leads to a similar simplification
in the representation of states, but in inhomogeneous models there can
be an additional complication for the volume operator.

\section{Classical phase space}

In the main part of this paper we are interested in spherical symmetry
where $S\cong{\rm SU}(2)$ (in general, the action on $P$ does not
project to an ${\rm SO}(3)$ action, even if it does so on $\Sigma$)
and, outside symmetry centers, $F\cong {\rm U}(1)$ such that $S/F\cong
S^2$. The reduced (radial) manifold $B$ is 1-dimensional. On the
orbits we have an invariant metric which can be written as
$\md\vt^2+\sin^2\vt\md\vp^2$ in angular coordinates which will be used
from now on. A coordinate on $B$ will be called $x$ in what follows,
but not fixed. The reduced phase space of this model has been studied
in ADM variables \cite{Kuchar} and complex Ashtekar variables
\cite{SphKl1,SphKl2}, which can be used for a reduced phase space or
Dirac quantization. Many relations in complex Ashtekar variables also
apply here, but one should be cautious since our notation is slightly
different and in some places adapted to a loop quantization.  The
classical model in real Ashtekar variables and preliminary steps of a
loop quantization have been described in \cite{SymmRed,PhD}.

Any invariant connection allowing a non-degenerate dual vector field
can be written as
\begin{equation} \label{A}
 A=A_x(x)\Lam_3\md r+(A_1(x)\Lam_1+A_2(x)\Lam_2)\md\vt+
(A_1(x)\Lam_2-A_2(x)\Lam_1)\sin\vt\md\vp+ \Lam_3\cos\vt\md\vp
\end{equation}
with three real functions $A_x$, $A_1$ and $A_2$ on $B$. The ${\rm
su}(2)$-matrices $\Lam_I$ are constant and are identical to
$\tau_I=-\frac{i}{2}\sigma_I$ or a rigid rotation thereof. An
invariant densitized triad has a dual form,
\begin{equation} \label{E}
 E=E^x(x)\Lam_3\sin\vt\frac{\partial}{\partial x}+
(E^1(x)\Lam_1+E^2(x)\Lam_2)\sin\vt\frac{\partial}{\partial\vt}+
(E^1(x)\Lam_2-E^2(x)\Lam_1)\frac{\partial}{\partial\vp}
\end{equation}
such that the functions $E^x$, $E^1$ and $E^2$ on $B$ are canonically
conjugate to $A_x$, $A_1$ and $A_2$:
\begin{equation}
 \Omega_{B}=\frac{1}{2\gamma G}\int_B\md x(\md
A_x\wedge\md E^x+ 2\md A_1\wedge\md E^1+2\md A_2\wedge\md E^2)
\end{equation}
with the gravitational constant $G$ and the Barbero--Immirzi parameter
$\gamma$.

It will later be useful to keep in mind a peculiarity of
one-dimensional models concerning the density weight of fields. As in
the full theory, the connection has density weight zero, and the
densitized triad is a vector field with density weight one. But in one
dimension the transformation properties with fixed orientation imply
that a 1-form is equivalent to a scalar of density weight one, while a
densitized vector field is equivalent to a scalar without density
weight. Under a coordinate change $x\mapsto y(x)$, a densitized vector
field, for instance, transforms as $E^a=\tilde{E}^b\partial
x^a/\partial y^b \abs{\det\partial y/\partial x}$, which implies
$E^x=\tilde{E}^x\cdot \abs{y'(x)}/y'(x)=\pm E^x$. Thus, $E^x$ can be
seen as the component of a densitized vector field on $B$ or as a
scalar, while $E^1$ and $E^2$ are densitized scalars (or 1-form
components). Similarly, $A_x$ is the component of a 1-form on $B$ or a
densitized scalar, while $A_1$ and $A_2$ are scalars (or densitized
vector field components).

These variables are subject to constraints which are obtained by
inserting the invariant forms into the full expressions. We have the
Gauss constraint
\begin{equation}
 G[\lambda]=\int_B\md x\lambda (E^x{}'+2A_1E^2-2A_2E^1)\approx0
\end{equation}
generating U(1)-gauge transformations, the diffeomorphism constraint
\begin{equation}
 D[N_x]=\int_B\md x N_x(2A_1'E^1+2A_2'E^2-A_xE^x{}')
\end{equation}
and the Euclidean part of the Hamiltonian constraint
\begin{eqnarray}
 H[N]&=&2\int_B\md x N \left(\abs{E^x}((E^1)^2+(E^2)^2)\right)^{-1/2} \\
&&\times\left(E^x(E^1A_2'-E^2A_1')+A_xE^x(A_1E^1+A_2E^2)+(A_1^2+A_2^2-1)
\nonumber
((E^1)^2+(E^2)^2)\right)\,.
\end{eqnarray}

In what follows it will be more convenient to work with variables that
are better adapted to the gauge transformations. We introduce the
gauge invariant quantities
\begin{eqnarray}
 A_{\vp}(x) &:=& \sqrt{A_1(x)^2+A_2(x)^2}\,,\\
 E_{\vp}(x) &:=& \sqrt{E^1(x)^2+E^2(x)^2}
\end{eqnarray}
and the internal directions
\begin{eqnarray}
 \Lambda_{\vp}^A(x) &:=& (A_1(x)\Lam_2-A_2(x)\Lam_1)/A_{\vp}(x)\,,\\
 \Lambda_{\vp}^E(x) &:=& (E^1(x)\Lam_2-E^2(x)\Lam_1)/E^{\vp}(x)
\end{eqnarray}
in the $\Lam_1$-$\Lam_2$ plane.  Furthermore, we parameterize
$\Lambda^A_{\vp}(x)$ and $\Lambda^E_{\vp}(x)$, which in general are
different from each other, by two angles $\alpha(x)$, $\beta(x)$:
\begin{eqnarray}
 \Lambda_{\vp}^A(x) &=:& \Lam_1\cos\beta(x)+\Lam_2\sin\beta(x)\,,\\
 \Lambda_{\vp}^E(x) &=:& \Lam_1\cos\left(\alpha(x)+\beta(x)\right)+
\Lam_2\sin\left(\alpha(x)+\beta(x)\right) \,.
\end{eqnarray}
Note that $\cos\alpha=\Lambda_{\vp}^A\cdot\Lambda_{\vp}^E$ is
gauge invariant under U(1)-rotations, while the angle $\beta$ is pure
gauge.

In these new variables the symplectic structure becomes
\begin{eqnarray}
 \Omega_{B} &=& \frac{1}{2\gamma G}\int_B\md x \left(\md
A_x\wedge\md E^x+2\md A_{\vp}\wedge\md (E^{\vp}\cos\alpha)+
2\md\beta\wedge \md(A_{\vp}E^{\vp}\sin\alpha)\right)\nonumber\\
 &=& \frac{1}{2\gamma G}\int_B\md x \left(\md A_x\wedge\md E^x+ \md
A_{\vp}\wedge\md P^{\vp}+ \md\beta\wedge\md P^{\beta}\right)
\end{eqnarray}
with new momenta
\begin{equation}
 P^{\vp}(x):=2E^{\vp}(x)\cos\alpha(x)
\end{equation}
conjugate to $A_{\vp}$ and
\begin{equation}
 P^{\beta}(x):=2A_{\vp}(x)E^{\vp}(x)\sin\alpha(x)=
A_{\vp}(x)P^{\vp}(x)\tan\alpha(x)
\end{equation}
conjugate to $\beta$. The Gauss constraint then takes the form
\begin{equation} \label{Gauss}
 G[\lambda]=\int_B\md x\lambda(E^x{}'-P^{\beta})\approx0
\end{equation}
which is easily solved by $P^{\beta}=E^x{}'$ while the function
$A_x+\beta'$ is manifestly gauge invariant.

Using these variables, the situation is different from that in the
full theory in that the momentum conjugate to the connection component
$A_{\vp}$ is not the triad component $E^{\vp}$, which together with
the momentum $E^x$ would directly determine the geometry
\begin{equation} \label{metric}
 \md s^2=E^x(x)^{-1}E^{\vp}(x)^2\md x^2+E^x(x)(\md\vt^2+\sin^2\vt\md\vp^2)\,.
\end{equation}
Instead, the momentum $P^{\vp}$ is related to $E^{\vp}$ through the
angle $\alpha$. This angle is a rather complicated function of the
variables, depending also on connection components: $\tan\alpha=
(A_{\vp}P^{\vp})^{-1}P^{\beta}$. This will complicate the quantum
geometry since the fluxes $P$ will be basic variables with simple
quantizations, while geometric operators like the volume operator will
be more complicated. In homogeneous models
\cite{HomCosmo,IsoCosmo}, on the other hand, this complication does
not appear since the Gauss constraint (\ref{Gauss}) with constant
$E^x$ implies $P^{\beta}=0$ and thus $\alpha=0$.

\section{Kinematical Hilbert space}

By definition, symmetric states can be described by restricting states
of the full theory to invariant connections, which are of the form
(\ref{A}) in the spherically symmetric case (from a different point of
view, focusing on coherent states, spherical symmetry has been
considered in \cite{BHCoh}). Using all states in the full theory this
leads to a complete, but not independent set of symmetric states,
which then must be functionals of $A_x(x)$, $A_1(x)$ and
$A_2(x)$. That such functionals have to be expected is also obvious
from the reduced point of view where one just quantizes the
classically reduced phase space. However, the class of functions
obtained in this way depends on the quantization procedure, a loop
quantization giving different results than, e.g., a Wheeler--DeWitt
like quantization (as happens already in the isotropic case where both
quantizations result in inequivalent representations \cite{Bohr}).

We first follow a reduced quantization point of view analogous to that
followed in \cite{Bohr}. In constructing the quantum theory we perform
the steps of the full loop quantization, thus obtaining a loop
quantization of the reduced model. Thereafter we will reduce states
from the full theory and implement the reduction there, leading to the
same results in particular for basic operators.

\subsection{Reduced quantization}

We start by choosing elementary functions on the classical
phase space that will be promoted to basic operators of the quantum
theory, acting on a suitable Hilbert space. The hallmark of loop
quantizations is that those basic quantities are chosen to be
holonomies of the connection and fluxes of the densitized triad. This
choice incorporates a smearing of the classical fields along lines
and surfaces, which is necessary for a well-defined representation,
and does so in a background independent manner. From the reduced point
of view, $A_{\vp}$ is a scalar for which there are analogous
techniques \cite{ScalarBohr,FermionHiggs} which we will use below.

\subsubsection{Cylindrical states}

A loop quantization in the connection representation is based on
cylindrical functions which depend on the connection only via
holonomies. If we just consider the space ${\cal A}_B$ of reduced
U(1)-connections given by $A_x(x)$ on $B$, cylindrical functions are
continuous functions on the space of generalized connections
$\bar{\cal A}_B$. As in the full theory, $\bar{\cal A}_B$ can be
written as a projective limit over graphs in $B$, which in the
1-dimensional case are simply characterized by a disjoint union of
non-overlapping edges, $g=\dot{\bigcup}_i e_i$, whose vertex set
$V(g)$ is the union of all endpoints of the $e_i$.  Choosing an
orientation of $B$, we fix the orientation of all edges to be
compatible with that of $B$. Holonomies then define spaces ${\cal
A}^g_B$ of maps from the set of edges of a given graph of $n$ edges to
${\rm SU}(2)^n$, which for classical connections reduces to $A_B\colon
g\to {\rm U(1)}, e\mapsto h^{(e)}:=\exp \frac{1}{2}i\int_e A_x(x)$
(the factor $1/2$ comes from taking matrix elements of
$\Lambda_3$-holonomies). The space of generalized connections is
obtained as the projective limit
\begin{equation}
 \bar{\cal A}_B=\varprojlim\nolimits_{g\subset B} {\cal A}^g_B
\end{equation}
with the usual projections $p_{gg'}(\bar{A}_B^g):=\bar{A}_B^g|_{g'}$
for $g'\subset g$.

Since $A_{\vp}$ transforms as a scalar, its holonomies with respect to
edges in $B$ would not be well-defined. Instead, following
\cite{FermionHiggs,ScalarBohr} one considers ``point holonomies''
$\exp(i\mu A_{\vp}(x))$ such that, for a fixed point $x\in B$, the
relevant space of states is the space $C(\bar{\R}_{\rm Bohr})$ of
continuous (almost periodic) functions on the Bohr compactification of
the real line $\R\ni A_{\vp}(x)$. The remaining independent scalar
function in the connection, $\beta(x)$, takes values in the circle
$S^1$ which is already compact. Corresponding point holonomies are
simply exponentials $\exp(i\beta(x))\in{\rm U}(1)$.

Since all points are independent, the space of generalized fields
$\bar{\cal A}_{S^2}$ is again a projective limit, this time over sets
of points $\{x_i\}_{i=1\ldots m}\subset B$, which can be taken as the
vertex set $V(g)$ of a graph $g$. For a fixed graph, we obtain the
space ${\cal A}^g_{S^2}$ of maps from the set $V(g)$ of $m$ vertices
to $(\bar{\R}_{\rm Bohr}\times {\rm U}(1))^m$, which for classical
fields is $A_{S^2}\colon V(g)\to \bar{\R}_{\rm Bohr}\times {\rm
U}(1),v\mapsto (A_{\vp}(v),e^{i\beta(v)})$. The space of generalized
fields is
\[
 \bar{\cal A}_{S^2}=\varprojlim\nolimits_{g\subset B} {\cal A}^g_{S^2}
\]
which can easily be combined with $\bar{\cal A}_B$ to obtain the space
of generalized spherically symmetric connections
\begin{equation}
 \bar{\cal A}_{B\times S^2}=\varprojlim\nolimits_{g\subset B} {\cal
A}^g_B\otimes {\cal A}^g_{S^2}\,.
\end{equation}

Since $\bar{\cal A}_{B\times S^2}$ is the projective limit of tensor
products of compact groups, U(1) and $\bar{\R}_{\rm Bohr}$, it carries
a normalized Haar measure which is analogous to the
Ashtekar--Lewandowski measure in the full theory and will be called
$\mu_0$. The kinematical Hilbert space is then obtained by completing
the space of cylindrical functions on $\bar{\cal A}_{B\times S^2}$
with respect to $\mu_0$. Holonomies defined above act by
multiplication on this space.

As in the full theory, one can use spin network states as a convenient
basis, which in the connection representation become functionals of
$A_x$, $A_{\vp}$ and $\beta$. They are cylindrical states based on a
given graph $g$ whose edges $e$ are labeled by irreducible
U(1)-representations $k_e\in\Z$, and whose vertices $v$ are labeled by
irreducible $\bar{\R}_{\rm Bohr}$-representations $\mu_v\in\R$ as well
as irreducible $S^1$-representation $k_v\in\Z$. The value of
such a spin network state in a given (generalized) spherically
symmetric connection $A$ then is
\begin{eqnarray} \label{SpinNetwork}
 T_{g,k,\mu}(A) &=& \prod_{e\in g} k_e(h^{(e)})\prod_{v\in
 V(g)}\mu_v(A_{\vp}(v)) k_v(\beta(v))\nonumber\\ &=& \prod_{e\in g}
 \exp\left(\tfrac{1}{2}ik_e\smallint_e A_x(x)\md x\right) \prod_{v\in
 V(g)} \exp(i\mu_v A_{\vp}(v)) \exp(ik_v\beta(v))\,.
\end{eqnarray}
Since $A_{\vp}$ and $\beta$ are scalars on $B$, they are not
integrated over in the states. On the other hand, $A_x$ as a
connection component is integrated to appear only via
holonomies. Alternatively, as discussed before, we can view the
one-dimensional connection component $A_x$ as a density-valued
scalar. Also from this perspective it would have to appear integrated
along regions in the above form. Since $A_{\vp}$ is by definition
non-negative, we will restrict the states to only those values.

\subsubsection{Flux operators}

For the flux of $E^x$ it is also helpful to view it in the
unconventional way as a scalar. At a given point $x$, $E^x(x)$ will
then simply be quantized to a single derivative operator without
integration:
\begin{equation}
 \hat{E}^x(x)f(h) = -i\frac{\gamma\lP^2}{4\pi}\sum_e\frac{\partial
 f}{\partial h^{(e)}} \frac{\delta h^{(e)}}{\delta A_x(x)}=
 \frac{\gamma\lP^2}{8\pi}\cdot\frac{1}{2}\sum_{e\ni x}h^{(e)}
\frac{\partial f}{\partial h^{(e)}}
\end{equation}
where $f$ is a cylindrical function depending on the holonomies
$h^{(e)}=\exp(\frac{1}{2}i\int_e A_x\md x)$. To simplify the notation
we assumed that $x$ lies only at boundary points of edges, which can
always be achieved by subdivision, and which contributes the
additional $\frac{1}{2}$. The other flux components, $P^{\vp}$ and
$P^{\beta}$, are density valued scalars and thus will be turned to
well-defined operators after integrating over regions ${\cal I}\subset
B$. We obtain
\begin{eqnarray}
 \int_{\cal I}\hat{P}^{\vp} f(h) &=& -i\frac{\gamma\lP^2}{4\pi} \int_{\cal
 I}\frac{\delta}{\delta A_{\vp}(x)}\md x f(h)=
 -i\frac{\gamma\lP^2}{4\pi} \int_{\cal I}\md x\sum_v \frac{\partial
 f}{\partial h^{(v)}} \frac{\delta h^{(v)}}{\delta A_{\vp}(x)}\nonumber\\
 &=& -i\frac{\gamma\lP^2}{4\pi} \sum_v\int_{\cal I}\md x \frac{\partial
 f}{\partial h^{(v)}}\delta(v,x) = -i\frac{\gamma\lP^2}{4\pi}
 \sum_{v\in{\cal I}} \frac{\partial}{\partial A_{\vp}(v)} f(h)
\end{eqnarray}
with $h^{(v)}:=A_{\vp}(v)$, and similarly
\begin{equation}
 \int_{\cal I}\hat{P}^{\beta} f(h)=
 -i\frac{\gamma\lP^2}{4\pi}\sum_{v\in{\cal I}}
\frac{\partial}{\partial \beta(v)} f(h)\,.
\end{equation}

Acting on spin network states (\ref{SpinNetwork}), we obtain
\begin{eqnarray}
 \hat{E}^x(x) T_{g,k,\mu} &=& \frac{\gamma\lP^2}{8\pi}
\frac{k_{e^+(x)}+k_{e^-(x)}}{2} T_{g,k,\mu} \label{Exspec}\\
 \int_{\cal I}\hat{P}^{\vp}T_{g,k,\mu} &=& \frac{\gamma\lP^2}{4\pi}
\sum_{v\in{\cal I}} \mu_v T_{g,k,\mu}\label{Ppspec}\\
 \int_{\cal I}\hat{P}^{\beta} T_{g,k,\mu} &=&
\frac{\gamma\lP^2}{4\pi}\sum_{v\in{\cal I}} k_v T_{g,k,\mu} \label{Pbspec}
\end{eqnarray}
where $e^{\pm}(x)$ are the two edges (or two parts of a single edge)
meeting in $x$. Thus, spin networks are eigenstates of all flux
operators and all flux operators have discrete spectra (normalizable
eigenstates). Note in particular that $\int_{\cal I}\hat{P}^{\vp}$ is
self-adjoint even though the range of $A_{\vp}$ is restricted to
non-negative values. In a Schr\"odinger representation the
corresponding derivative operator would not have a self-adjoint
extension, which is the case here on the restricted Bohr Hilbert
space.

Knowing the flux operators allows us to quantize and solve the Gauss
constraint (\ref{Gauss}). Restricting attention for simplicity to
piecewise constant $\lambda$, it suffices to quantize the integrated
density $E^x{}'$, which can be done easily by using $\int_{x_-}^{x_+}
E^x{}'\md x=E^x(x_+)-E^x(x_-)$. Thus,
\begin{equation}
 \hat{G}[\lambda] T_{g,k,\mu} = \frac{\gamma\lP^2}{8\pi}\sum_v \lambda(v)
(k_{e^+(v)}-k_{e^-(v)}-2k_v) T_{g,k,\mu} = 0
\end{equation}
which is solved by
\begin{equation} \label{kGauss}
 k_v=\tfrac{1}{2}(k_{e^+(v)}-k_{e^-(v)})
\end{equation}
for all vertices $v$ of a given spin network. Since the $k_v$ must be
integer, all differences $k_{e^+(v)}-k_{e^-(v)}$ must be even,
restricting the allowed values.  The labels $k_v$ are then determined
completely by the edge labels $k_e$ and can be dropped while the other
vertex labels, $\mu_v$, are not restricted by the Gauss constraint. In
fact, when (\ref{kGauss}) is satisfied, the spin network
(\ref{SpinNetwork}) takes the form
\begin{equation} \label{GaugeInvSpinNetwork}
 T_{g,k,\mu}=\prod_e \exp\left(\tfrac{1}{2}i k_e
\smallint_e(A_x+\beta')\md x\right)  \prod_v
\exp(i\mu_v A_{\vp(v)})
\end{equation}
which depends only on manifestly gauge invariant quantities.

The diffeomorphism constraint only generates transformations along
$B$, which can be dealt with by group averaging in complete analogy to
the full theory (see also \cite{SymmRed,PhD} for more details).

\subsection{Spherically symmetric states from the full theory}

Since symmetric states are by definition full states supported on
invariant connections, one can, for a given symmetry action, find the
form of symmetric states by restricting a complete set of full
states. In a second step, one can than select an independent set of
the resulting functions on $\bar{\cal A}_{B\times S^2}$ and
orthonormalize it in the Haar measure. Gauge transformations in the
symmetric context are usually more restricted than those in the full
theory; for instance along homogeneous directions gauge
transformations are required to be constant in order to preserve the
form of invariant connections like (\ref{A}). Therefore, one can and
should also consider certain gauge non-invariant states in the full
theory in order to access all possible (gauge invariant) states in the
model.

\subsubsection{States}

We start with spin network states in the full theory which for
simplicity will be assumed to be based on graphs made of only radial
edges, i.e.\ submanifolds of $B\times\{p\}$ where $p\in S^2$ is fixed,
or edges lying in orbits of the rotation group. In the latter case we
assume that the edge is composed of edges along great circles in the
$\vt$-direction or at $\vt=\pi/2$ (the last restriction allows us
to ignore the field independent term in the connection which
corresponds to a component of the spin connection). Note that the
composition of orbital edges does not need to be a closed edge, which
still corresponds to a gauge invariant state from the reduced point of
view. For orbital edges we use the angular coordinates as parameters,
while there is no preferred parameterization for radial
edges. Furthermore, after fixing an orientation of the radial manifold
$B$ we choose all radial edges to be oriented in the same way as
$B$. The set of states based on those graphs is certainly not complete
in the full theory, but it will be sufficient for the spherically
symmetric sector. In particular, those states suffice to separate
spherically symmetric connections. At this point states are not
simplified much since vertices can still have arbitrary valence.

Let us thus fix such a state, based on a certain graph of the above
form. Its reduction is performed by inserting holonomies obtained from
(\ref{A}). We then need only two parameters for each orbital edge, the
coordinate length $\mu$ of the edge and the position $x$ of the orbit
along $B$. This leads to holonomies
\begin{eqnarray} \label{hols}
 h^{e}_x(A) &=& \exp\int_e A_x(x)\md x \Lam_3 = \cos
\tfrac{1}{2}\int_e A_x(x)\md
x+ 2\Lam_3\sin\tfrac{1}{2}\int_e A_x(x)\md x\label{holx}\\
 h^{(x,\mu_{\vt})}_{\vt}(A) &=& \exp\smallint_0^{\mu_{\vt}}A_{\vp}(x)
\md\vt\Lambda^A_{\vt}(x)= 
\cos\tfrac{1}{2}\mu_{\vt} A_{\vp}(x)+
2\Lambda^A_{\vt}(x)\sin\tfrac{1}{2}\mu_{\vt} A_{\vp}(x)\nonumber\\
&=& \cos\tfrac{1}{2}\mu_{\vt} A_{\vp}(x)+
2e^{-\pi\Lam_3/2}\Lambda^A_{\vp}(x)e^{\pi\Lam_3/2}
\sin\tfrac{1}{2}\mu_{\vt} A_{\vp}(x) \label{holvt} \\
 h^{(x,\mu_{\vp})}_{\vp}(A) &=& \exp\left(-\smallint_0^{\mu_{\vp}}
\md\vp A_{\vp}(x) \Lambda^A_{\vp}(x)\right)= 
\cos\tfrac{1}{2}\mu_{\vp} A_{\vp}(x)-
2\Lambda_{\vp}^A(x)\sin\tfrac{1}{2}\mu_{\vp} A_{\vp}(x)
\end{eqnarray}
where path orderings are not necessary since $\Lam_3$ is constant
along $B$, and the $\Lambda^A(x)$ are constant along paths on orbits.

The parameters $\mu_{\vt}$ and $\mu_{\vp}$ are simply the parameter
lengths of the orbital edges in the $\vt$- and $\vp$-directions. They
can take any real value since we do not require that individual
orbital edges run around great circles once, but can also run through
just part of a great circle, or also through the same great circle
several times in both directions.

Inserting these holonomies is most easily done for an alternative
form of the states rather than the spin network basis. All states can
be obtained (in an overcomplete way) as products of Wilson loops which
in our case are composed of radial or orbital holonomies. (If $B$ has
a boundary, there can be open ends of the loop where gauge
transformations are frozen.) Each such state is a superposition of
matrix elements of the form
\begin{eqnarray*}
 && h_x^{(e_1)}\cdot h_{\vt}^{(v_1,\mu^1_{\vt,1})}
 h_{\vp}^{(v_1,\mu^1_{\vp,1})} \cdots h_{\vt}^{(v_1,\mu^1_{\vt,n_1})}
 h_{\vp}^{(v_1,\mu^1_{\vp,n_1})}\\
&&\cdot h_x^{(e_2)}\cdot
 h_{\vt}^{(v_2,\mu^2_{\vt,1})} h_{\vp}^{(v_2,\mu^2_{\vp,1})} \cdots
 h_{\vt}^{(v_2,\mu^2_{\vt,n_2})} h_{\vp}^{(v_2,\mu^2_{\vp,n_2})}\\
&& \cdot
 h_x^{(e_3)} \cdots
\end{eqnarray*}
with radial edges $e_1$, $e_2$, $e_3$, \ldots, and vertices $v_1$,
$v_2$, \ldots, where $v_1$ is the endpoint of $e_1$ and the starting
point of $e_2$. (A given edge $e$ can appear several times in such an
expression since it can be traversed back and forth with running
through orbital edges in between.) The parameters $\mu^i_{\vt/\vp,j}$
are the parameter lengths of orbital edges and can take any real
value.

To simplify the general expressions evaluated in spherically symmetric
connections we assume that the states are gauge invariant under gauge
transformations around $\Lam_3$, constant on the orbits (which still
allows also open graphs). We can then gauge the angle $\beta(x)$ to be
constant (with a local gauge transformation $\exp(-\beta(x)\Lam_3)$,
possibly up to a global gauge transformation if there is a
boundary). Then also $\Lambda_{\vp}^A$ is constant and we can apply
Lemma \ref{lemma} to (almost) commute the holonomies. In particular,
we can order the radial holonomies according to coordinate values $x$
of their starting points, also re-orienting them if necessary such
that they run in the positive orientation of $B$. Between different
edges there are vertices which can have the following forms:
\begin{eqnarray} \label{reduce}
 &&\cdots (h_x^{(e_-)})^{k_-}\cdot h_{\vp}^{(v,\mu)}\cdot (h_x^{(e_+)})^{k_+}
\cdots, \nonumber\\
 &&\cdots (h_x^{(e_-)})^{k_-} \Lam_3\cdot h_{\vp}^{(v,\mu)}\cdot
(h_x^{(e_+)})^{k_+} \cdots\quad\mbox{or} \nonumber\\
 &&\cdots (h_x^{(e_-)})^{k_-}\cdot h_{\vp}^{(v,\mu)}\cdot \Lam_3
(h_x^{(e_+)})^{k_+}\cdots 
\end{eqnarray}
where possible factors of $\Lam_3$ come from
$\exp(\frac{1}{2}\pi\Lam_3)$ in $\vt$-holonomies (\ref{holvt}).

Matrix elements of the resulting products of holonomies can easily be
seen to be superpositions of states of the form
(\ref{GaugeInvSpinNetwork}), keeping in mind that we chose the gauge
such that $\beta$ is constant along $B$. It is also possible, though
more tedious, to follow this procedure without fixing the gauge. We
just mention the example of a single ``rectangular'' loop made of one
radial edge with holonomy $h_x$ and two orbital ones along the equator
at $x_{\pm}$ with holonomies $h_{\pm}$. The corresponding Wilson loop
is
\begin{eqnarray} \label{square}
 \tr(h_xh_+h_x^{-1}h_-^{-1}) &=& \tr\left( \left(\cos\tfrac{1}{2}\smallint
A_x+2\Lam_3\sin\tfrac{1}{2}\smallint 
A_x\right) (\cos \tfrac{1}{2}A_{\vp}(x_+)+2\Lam^A_{\vp}(x_+)\sin
\tfrac{1}{2}A_{\vp}(x_+))\right. \nonumber\\ 
 && \left.\left(\cos\tfrac{1}{2}\smallint
A_x-2\Lam_3\sin\tfrac{1}{2}\smallint A_x\right) (\cos 
\tfrac{1}{2}A_{\vp}(x_-)-2\Lam^A_{\vp}(x_-)\sin
\tfrac{1}{2}A_{\vp}(x_-))\right)\nonumber\\ 
 &=& 2\cos \tfrac{1}{2}A_{\vp}(x_+) \cos
\tfrac{1}{2}A_{\vp}(x_-)\nonumber\\
 && +2
\cos\left(\smallint A_x+\beta(x_+)-\beta(x_-)\right) \sin
\tfrac{1}{2}A_{\vp}(x_+) \sin \tfrac{1}{2}A_{\vp}(x_-)
\end{eqnarray}
where we used $\tr(\Lambda_{\vp}^A(x_+)\Lambda_{\vp}^A(x_-))=
-\frac{1}{2}\cos(\beta(x_+)-\beta(x_-))$ and
$\tr(\Lam_3\Lambda_{\vp}^A(x_+)\Lambda_{\vp}^A(x_-))=
\frac{1}{4}\sin(\beta(x_+)-\beta(x_-))$. This state can clearly be
written as a superposition of states (\ref{GaugeInvSpinNetwork}).

Thus, the states obtained before from a loop quantization of the
classically reduced phase space also emerge as symmetric states in the
full theory. This is true, however, only with a slight restriction
since full gauge invariant spin network states evaluated in
spherically symmetric connections satisfy an additional condition: the
gauge transformation $\exp(\frac{1}{2}\pi\Lam_3)$ changes the sign of
$A_1$ and $A_2$ everywhere, which means that all those states will be
even under changing the sign of all $A_{\vp}(x)$, as e.g.\
(\ref{square}). This can easily be imposed as an additional condition
on the reduced states, and it will be respected by operators coming
from full ones.

\subsubsection{Flux operators}

In general, the dual action of operators of the full theory applied to
distributional symmetric states will not lead to another symmetric
state. The reason is that symmetric states only incorporate the
condition for the connection to be invariant, but if full operators
are used there is no condition for an invariant triad. In such a case,
even classically the Hamiltonian flow generated by an arbitrary
function on the phase space would in general leave the subspace of
invariant connections with arbitrary triads (while the flow would
always stay inside the subspace of invariant connections and invariant
triads if the symmetric model is well-defined). There are, however,
notable exceptions which allow us to obtain all operators for the
basic variables directly from the full theory. This is true for
holonomies of $A_x$ and $A_{\vp}$ which commute with connections,
anyway. But we can also find special fluxes whose classical expressions
generate a flow that stays in the subspace of invariant
connections. For instance, for the $\Lam_3$-component of a full flux
for a symmetry orbit $S^2$, $F_{S^2}^3(x):=\int_{S^2}\Lambda_3\cdot
(E(x)\lrcorner\md x)\md^2y$, we have
\[
 \{A_a^i(x),F_{S^2}^3\}|_{{\cal A}_{\rm inv}\times{\cal
E}}=\gamma\kappa\Lambda_3^i\delta_a^x \int_{S^2}\delta(x,y)\md^2y
\]
which defines a distributional vector field on the phase space
parallel to the subspace ${\cal A}_{\rm inv}\times{\cal E}$ of
invariant connections (parallel to $A_x$). If we would look at any
other internal component, e.g., $F_{S^2}^2$ using $\Lam_2$, on the
other hand, the Poisson bracket would be proportional to
$\Lam_2^i\delta_a^x$, which is not parallel to the subspace of
invariant connections. Similarly, one can see that the flux
\[
 F_{{\cal
 I}\times S^1}:=\int_{{\cal I}\times S^1}
 \left(\Lambda_{\vp}^A(x)\cdot (E(x)\lrcorner\md\vp) \md
x\md\vt+\Lambda_{\vt}^A(x)\cdot (E(x)\lrcorner\md\vt) \md
x\md\vp\right)
\]
for a
cylindrical surface along an interval ${\cal I}\subset B$ generates a
flow parallel to $A_{\vp}$, which leaves the space of invariant
connections invariant.

These two fluxes are sufficient for the basic momenta since
\[
 \int_{S^2}\Lam_3\cdot (E(x)\lrcorner\md
 x)\md^2y= 4\pi E^x(x)
\]
and 
\[
 \int_{{\cal I}\times S^1} \left(\Lambda_{\vp}^A(x)\cdot
 (E(x)\lrcorner\md\vp) \md x\md\vt+\Lambda_{\vt}^A(x)\cdot
 (E(x)\lrcorner\md\vt) \md x\md\vp\right)= 4\pi
\int_{\cal I} P^{\vp}(x)\md x
\]
whose quantizations can thus be obtained directly from the full
theory. Note that this would not be possible for $E^{\vp}$, for
instance, since its corresponding flux would generate a transformation
changing the invariant form of $A$ (since any full expression reducing
to $E^{\vp}$ upon reduction would involve the non-linear function
$\Lambda_E^{\vp}$ depending on the triad components).

For the flow $F_{S^2}^3(x)$ we obtain the $\Lam_3$-component of an
invariant vector field associated with the edges containing $x$. The
pull back to invariant connections in (\ref{dist}) ensures that the
dual action on the distribution $\Psi$ can be expressed by an
invariant vector field on the representation $\psi$ where only radial
holonomies $h_x^{(e)}$ of the form (\ref{holx}) appear. For the
explicit expression we again assume that $x$ is an endpoint of two
edges, $e^+(x)$ and $e^-(x)$ which can be achieved by appropriate
subdivision, and obtain
\begin{equation}
 \hat{F}_{S^2}^3(x) = \tfrac{1}{2}i\gamma\lP^2 \left(\tr(\Lam_3
h_x^{(e^+(x))})^T \frac{\partial}{\partial h_x^{(e^+(x))}} + \tr(\Lam_3
h_x^{(e^-(x))})^T \frac{\partial}{\partial h_x^{(e^-(x))}} \right)\,.
\end{equation}
Since $\Lam_3$ commutes with radial holonomies $h_x^{(e)}$, we do not
need to distinguish between left and right invariant vector field
operators. According to the derivation of states above, they can be
seen as polynomials in the radial holonomies. The action of a
derivative operator $\tr(\Lam_3 h_x^{(e)})^T
\partial/\partial h_x^{(e)}$ with respect to $h_x^{(e)}$
then amounts to replacing $(h_x^{(e)})^k$ by $k \Lam_3 (h_x^{(e)})^k$
and $\Lam_3 (h_x^{(e)})^k$ by $-\frac{1}{4}k (h_x^{(e)})^k$ (note that
insertions of $\Lam_3$ appear automatically as in (\ref{reduce}); in
any case, they would occur when considering a more general class of
gauge non-invariant states in the full theory which are invariant from
the reduced point of view). Eigenstates of the derivative can then be
obtained by forming linear combinations such that only the
combinations $(h_x^{(e)})^k\pm 2i\Lam_3(h_x^{(e)})^k$ appear, which
are mapped to $\mp \frac{1}{2}ik((h_x^{(e)})^k\pm
2i\Lam_3(h_x^{(e)})^k)$. In this way, one obtains eigenstates of
$\hat{E}^x(x)= (4\pi)^{-1}\hat{F}_{S^2}^3(x)$ and a spectrum identical
to (\ref{Exspec}).

The operator $\hat{E}^x$ also appears in the Gauss constraint. A gauge
invariant state in the full theory in particular satisfies
$\Lam_3\cdot(J_L(h_x^{(e^+(x))})- J_R(h_x^{(e^-(x))})+
J_L(h_{\vp}^{(x)})- J_R(h_{\vp}^{(x)}))=0$ in any vertex $x$ where we
can assume the form (\ref{reduce}) for the spin network (a general
vertex would just be a superposition of those vertices). The operators
$\Lam_3\cdot J(h_x^{(e^{\pm}(x))})$ simply give operators $\hat{E}^x$
where we do not need to distinguish between right and left invariant
ones, while for the derivative operators with respect to
$\vp$-holonomies we have
\begin{eqnarray*}
 \Lam_3\cdot (J_L(h_{\vp}^{(x)})- J_R(h_{\vp}^{(x)}))&=&
-i\left(\tr(h_{\vp}^{(x)}\Lam_3)^T\frac{\partial}{\partial
h_{\vp}^{(x)}}- \tr(\Lam_3h_{\vp}^{(x)})^T\frac{\partial}{\partial
h_{\vp}^{(x)}}\right)\\
 &=& i\tr[\Lam_3,h_{\vp}^{(x)}]^T\frac{\partial}{\partial
h_{\vp}^{(x)}}= i\tr\frac{\partial h_{\vp}^{(x)}}{\partial \beta(x)}
\frac{\partial}{\partial h_{\vp}^{(x)}}\\
&=& i\frac{\partial}{\partial\beta(x)}
\end{eqnarray*}
using $[\Lam_3,h_{\vp}^{(x)}]= \partial
h_{\vp}^{(x)}/\partial\beta(x)$ with
$\Lambda^A_{\vp}(x)=\cos\beta(x)\Lam_1+\sin\beta(x)\Lam_2$. The right
hand side is then simply proportional to $\hat{P}^{\beta}(x)$ and we
see that the $\Lam_3$-component of the full quantum Gauss constraint
is identical to the reduced Gauss constraint. (The remaining
components of the full Gauss constraint would not fix the space of
symmetric states and thus cannot be dealt with in this way. They would
have to be satisfied identically.) In what follows we use the above
equation to define the operator $\hat{P}^{\beta}$ which then has the
spectrum (\ref{Pbspec}) as in the reduced case.

It remains to look at the full quantization of
\[
 \int_{\cal
 I}P^{\vp}\md x=\frac{1}{4\pi}\int_{{\cal I}\times S^1}
 \left(\Lambda_{\vp}^A\cdot(E\lrcorner\md\vp) \md x\md\vt+
 \Lambda_{\vt}^A\cdot(E\lrcorner\md\vt) \md x\md\vp\right)
\]
acting on symmetric states. Since they are now
$\Lambda_{\vp}^A$-components of derivative operators with respect to
$h_{\vp}$, the end result is again a simple derivative operator acting
on powers $h_{\vp}^{\mu}$, which gives a spectrum proportional to that
in (\ref{Ppspec}). It is only proportional since we chose the surface
for this flux using the great circle $S^1$, which contributes a factor
$2\pi$. Choosing other circles on the orbits would change the factor
which, anyway, can always be absorbed by a unitary
transformation. (Such a rescaling is unitary for this operator, as in
the isotropic case \cite{Bohr}, since the range of eigenvalues $\mu$
is the real line.)

\subsection{Reduced states from the full point of view}

The derivations presented above demonstrate that the states obtained
from the purely reduced point of view are also necessary in this form
when viewing them as being obtained by restricting full states. Also
basic flux operators obtained in the reduced quantization and from the
full theory via the dual action on distributional symmetric states
agree. Thus, the model can be seen as a symmetric sector of the full
theory, associated with a subspace of the distribution space ${\rm
Cyl}^*$. 

Full symmetric states which satisfy the 3-component of the full Gauss
constraint are also gauge invariant from the reduced point of view. In
fact, we can directly take the dual action of the 3-component of the
Gauss constraint to obtain the reduced constraint (the other
components do not map the space of symmetric states to itself). In the
reduced model, this has as a consequence that the difference of labels
associated with neighboring radial edges has to be even. The analog in
the full theory can be seen by considering vertices where radial and
orbital edges meet. After inserting invariant connections, a state
with such a vertex is equivalent to a superposition of states with a
vertex having one incoming radial edge, a composition of several
orbital edges, and an outgoing radial edge. From this point of view,
this gauge invariant vertex is a $2(n+1)$-vertex with the ingoing and
outgoing radial edges with representations $j_-$ and $j_+$,
respectively, as well as $n$ closed orbital edges which each
contribute one incoming and one outgoing part with spin $j_i$. For an
intertwiner we can first construct the tensor product of the orbital
representations, $\bigotimes_{i=1}^n j_i\otimes j_i= \bigoplus_i l_i$
where only integer $l_i$ occur in the decomposition. The vertex
intertwiner maps this representation to the tensor product $j_+\otimes
j_-$, which for integer $l_i$ is possible in a non-trivial way only if
$j_+$ and $j_-$ are either both integer or both $1/2$ times an
integer. Thus, $j_+-j_-\in\Z$, which is equivalent to the fact that
the difference of charges $k_+-k_-$ must be even.

Similarly, the reduced diffeomorphism constraint can be obtained
directly from the dual action of the full diffeomorphism constraint
for a radial shift vector. For other shifts, the dual action would not
fix the space of symmetric states. For a radial shift, then, the
constraint generates transformations which move vertices along the
radial manifold, which is the same as the action generated by the
reduced diffeomorphism constraint. Thus, also the reduced
diffeomorphism constraint can be obtained via the dual action on
symmetric states which leads to the same results as quantizing the
classically reduced constraint. The Hamiltonian constraint, on the
other hand, is non-linear in the triads such that its dual action
cannot be used.

We thus have seen how states and basic operators of the reduced model
can be obtained from the full theory. Composite operators can then be
built from the basic ones within the model. An analogous derivation of
composite reduced operators from those in the full theory is more
complicated since a direct application of the dual action would not
fix the space of symmetric states.

It follows from these considerations that the representation of the
reduced model is determined by that of the full theory. Since the
diffeomorphism covariant holonomy/flux representation of full loop
quantum gravity is unique under certain weak conditions
\cite{LOST}, a representation for
the model is selected naturally. Starting from the classically reduced
model, on the other hand, would have left open the choice of
representation. In such a case, the representation is usually selected
in such a way that explicit calculations are possible, which does not
say anything about physical correctness. Even working in the framework
of this paper and using the same variables, there would be other
possibilities. For instance, viewing the scalar density $P^{\vp}$ as a
1-form, which in one dimension has the same transformation properties
if the orientation is preserved, suggests to quantize it via
holonomies. In this case, $A_{\vp}$ rather than $P^{\vp}$ would become
discrete. Similarly, we could use point holonomies for the densitized
vector field component $E^x$, which can also be viewed as a
scalar. This would give a discrete $A_x$. All these alternatives are
possible only in the reduced model due to the special behavior under
coordinate transformations. But they are not possible in the full
theory and thus cannot be obtained when the link between the model and
the full quantization is taken into account. Studying these
representations further can shed light on physical properties and
effects that are unique to the loop representation of the full theory.

\subsection{Semiclassical geometry}

The flux eigenvalues allow us to find conditions for states which
would be expected in regimes where the spatial geometry is almost
classical. Comparing (\ref{metric}) with the Schwarzschild solution at
large radius, or general asymptotic conditions, shows that $E^x$
(corresponding to $r^2$ for Schwarzschild) should be large together
with $E^x{}'=P^{\beta}$. This implies that the edge labels $k_e$ and
the differences $k_{e^+(v)}-k_{e^-(v)}=2k_v$ have to be large compared
to one since eigenvalues of $\hat{E}^x$ and $\hat{P}^{\beta}$ are
directly given by the labels without summing over vertices. This is
analogous to the homogeneous case where for a semiclassical geometry
all labels have to be large.

The situation is different for the other triad component. From the
metric we see that also $E^{\vp}$ has to be large which, for generic
$\alpha$ implies that $P^{\vp}$ must be large. However, the
quantization of the density $P^{\vp}$ is well-defined only if it is
first integrated over an interval in $B$, which means that the
relevant eigenvalues are given by a vertex sum $\sum_{v}\mu_v$,
which needs to be large. This can be realized by large individual
$\mu_v$, or by a dense distribution of vertices such that many small
$\mu_v$ add up to a large value. This situation is analogous to that
in the full theory where geometric operators are always given by
vertex sums. It is then expected that states with many small labels
are relevant for a semiclassical geometry since they dominate the
counting of states.

At this point, the difference between $P^{\vp}$ and $E^{\vp}$ suggests
possible consequences for semiclassical physics. If we fix a
$\mu_{\vp}$ (which happens, e.g., if we consider the dynamics
\cite{Bohr}) and restrict the operators to a separable subspace of our
Hilbert space generated by $e^{i\mu_{\vp} A_{\vp}}$, we would have a
discrete set $\mu_{\vp}n$ of eigenvalues with integer $n$. Then, the
sum $\sum_v\mu_v=\mu_{\vp}\sum_v n_v$ would still be of the same form
and not become denser at large eigenvalues (as would happen in the
full theory for, e.g., the area operator). The triad component
$E^{\vp}$, which appears in the metric (\ref{metric}), however, is a
more complicated function of the basic variables and thus is likely to
have a more complicated vertex contribution leading to
crowded eigenvalues \cite{SphSymmVol}.

\section{Other $1+1$ models}

There are many other models in $1+1$ dimensions which have infinitely
many physical degrees of freedom, but are integrable
\cite{K:EinsteinRosen,Integrable}, and which would be interesting to
compare with a loop quantization. The general form of an invariant
connection in those cases is (\ref{Agen}) where the
$\Lambda_I(x)\in{\rm su}(2)$, $\tr(\Lambda_I(x)^2)=-\frac{1}{2}$ can
be restricted further depending on the symmetry action. In general,
however, they do not satisfy $\tr(\Lambda_I\Lambda_J)=
-\frac{1}{2}\delta_{IJ}$, which was the case in the spherically
symmetric model with its non-trivial isotropy group and was
responsible for the simplified structure of states and basic
operators.

In cylindrically symmetric models with a space manifold
$\Sigma=\R\times(S^1\times\R)$, for instance, the symmetry group
$S=S^1\times\R$ acts freely, and invariant connections and triads have
the form
\begin{eqnarray}
 A &=& A_x(x)\Lam_3\md x+ (A_1(x)\Lam_1+A_2(x)\Lam_2)\md z+
(A_3(x)\Lam_1+A_4(x)\Lam_2)\md\vp \\
 E &=& E^x(x)\Lam_3\frac{\partial}{\partial x}+
(E^1(x)\Lam_1+E^2(x)\Lam_2)\frac{\partial}{\partial z}+
(E^3(x)\Lam_1+E^4(x)\Lam_2)\frac{\partial}{\partial\vp}
\end{eqnarray}
such that $\tr(\Lam_3\Lambda_z)=0=\tr(\Lam_3\Lambda_{\vp})$, but in
general $\tr(\Lambda_z\Lambda_{\vp})\not=0$.

The corresponding metric is
\begin{eqnarray}
 \md s^2 &=& (E^x)^{-1}(E^1E^4-E^2E^3)\md x^2+ E^x(E^1E^4-E^2E^3)^{-1}
\left(((E^3)^2+(E^4)^2)\md z^2\right.\nonumber\\
 &&
- \left.(E^2E^4+E^1E^3)\md z\md\vp+ ((E^1)^2+(E^2)^2)\md\vp^2\right)
\end{eqnarray}
which is not diagonal. To simplify the model further one often
requires that the metric is diagonal, which physically corresponds to
selecting a particular polarization of Einstein--Rosen waves. This is
achieved by imposing the additional condition $E^2E^4+E^1E^3=0$ which,
in order to yield a non-degenerate symplectic structure, has to be
accompanied by $A_2A_4+A_1A_3=0$ for the connection components. Thus,
polarized cylindrical waves of this form also have perpendicular
internal directions since now $\tr(\Lambda_z\Lambda_{\vp})=0$ for both
$A$ and $E$, and similar simplifications as in the spherically
symmetric case can be expected. 

The form of the metric now is
\begin{equation}\label{metricER}
 \md s^2= (E^x)^{-1}E^zE^{\vp}\md x^2+ E^x \left(E^{\vp}/E^z\md z^2+
E^z/E^{\vp}\md\vp^2\right)
\end{equation}
with
\begin{equation}
 E^z:=\sqrt{(E^1)^2+(E^2)^2} \quad, \quad
E^{\vp}:=\sqrt{(E^3)^2+(E^4)^2}\,.
\end{equation}
Einstein--Rosen waves are usually represented in the form
\begin{equation} \label{metricERpsi}
 \md s^2 = e^{2(\gamma-\psi)}\md r^2+ e^{2\psi}\md z^2+
e^{-2\psi}r^2\md\vp^2
\end{equation}
with only two free functions $\gamma$ and $\psi$. Thus, compared with
(\ref{metricER}) one function has been eliminated by gauge fixing the
diffeomorphism constraint.

In fact, this form can be obtained from the more general
(\ref{metricER}) by a field-dependent coordinate change
\cite{BicakSchmidt}: The symmetry reduction leads to a space-time
metric $\md s^2= e^{\Lambda}\md U\md V+W(e^{-\Psi}\md x^2+e^{\Psi}\md
y^2)$ which indeed has a spatial part as in (\ref{metricER}) with
three independent functions $\Lambda$, $W=E^x$ and
$\Psi=\log(E^{\vp}/E^z)$. One then introduces $t:=\frac{1}{2}(V-U)$
and $\rho:=\frac{1}{2}(V+U)$ such that $\md s^2=e^{\Lambda}(-\md
t^2+\md\rho^2)+\rho(e^{-\Psi}\md x^2+e^{\Psi}\md y^2)$. Finally,
defining $\Lambda=2(\gamma-\psi)$, $e^{-2\psi}\rho:=e^{-\Psi}$ and
renaming $x=:\vp$, $y=:z$ leads to the metric
(\ref{metricERpsi}). Einstein's field equations then imply that $\psi$
behaves as a free scalar on a flat space-time, which can be quantized
with standard Fock techniques \cite{K:EinsteinRosen}. However, to
arrive at this form of the metric, several coordinate transformations
have been performed which mix coordinates with the physical
fields. This potentially eliminates any contact the model may have
with a full theory and indicates that results may be very particular
to this kind of model. From the point of view taken here, where the
quantum representation comes directly from the full theory, a
subsequent transformation in such a way is impossible, which also
means that the quantization of the model will be more complicated. We
will see soon that the transversal geometry given by (\ref{metricER})
becomes discrete after loop quantizing, which $\psi$ when treated as
an ordinary scalar will not be. Thus, the quantum geometries obtained
from both representations differ from each other, which can also lead
to differing physical results (as in the homogeneous case, where loop
properties are extremely different from Wheeler--DeWitt results
concerning the issue of singularities and also phenomenology). This
may in particular be of interest in view of large quantum gravity
effects derived from wave models \cite{LargeEff}.
On the other hand, a direct comparison between different quantizations
is made more complicated by the coordinate dependent field
transformation.

We now present the initial steps of a loop quantization along the
lines followed in the spherically symmetric model. This will allow us
to see some properties of the quantum geometry. Most of the steps to
arrive at the kinematical Hilbert space and basic operators can be
done almost identically to those followed before.

In analogy to the spherically symmetric model we now introduce
\begin{eqnarray}
 A_z:=\sqrt{A_1^2+A_2^2} \quad &,&\quad A_{\vp}:=\sqrt{A_3^2+A_4^2}\\
 \Lambda_z^A:=\frac{A_1\Lam_1+A_2\Lam_2}{A_z} \quad &,& \quad
\Lambda_{\vp}^A:=\frac{A_3\Lam_1+A_4\Lam_2}{A_{\vp}}
\end{eqnarray}
and analogously $E^z$, $E^{\vp}$, $\Lambda_E^z$ and
$\Lambda_E^{\vp}$. Furthermore, we write
\begin{eqnarray}
 \Lambda_{\vp}^A &=& \cos\beta\Lam_1+\sin\beta\Lam_2\\
 \Lambda_E^{\vp} &=&
\cos(\alpha+\beta)\Lam_1+\sin(\alpha+\beta)\Lam_2\,.
\end{eqnarray}
With the polarization condition, this implies
\begin{eqnarray*}
 \Lambda_z^A &=& -\sin\beta\Lam_1+\cos\beta\Lam_2\\
 \Lambda_E^z &=& -\sin(\alpha+\beta)\Lam_1+\cos(\alpha+\beta)\Lam_2
\end{eqnarray*}
such that we have only two angles, $\beta$ which is pure gauge and
$\alpha$ as in the spherically symmetric model.

The symplectic structure tells us that momenta of $A_z$ and $A_{\vp}$
are not given by triad components directly, but by
$P^z:=E^z\cos\alpha$ and $P^{\vp}:=E^{\vp}\cos\alpha$. The momentum of
$\beta$ is $P^{\beta}:=(A_zE^z+A_{\vp}E^{\vp})\sin\alpha$, which is
related to $E^x$ by the same Gauss constraint as in the spherically
symmetric case.

The adaptation of the construction of states and operators to this
model is now straightforward, the only difference being that we have
one additional degree of freedom per point on $B$, given by $A_z$ for
which we have additional holonomies $\exp(i\mu_z A_z)$ in vertices of
spin network states. As before, flux operators do not give us direct
information about the geometry since fluxes are related to the triad
in a more complicated way. Still, the orbital components of the metric
in the $z$ and $\vp$ directions are easily accessible since
$E^z/E^{\vp}=P^z/P^{\vp}$ thanks to a cancellation of
$\cos\alpha$. Thus, the spectrum of the orbital geometry can easily be
computed, after using techniques as in \cite{InvScale} to quantize the
inverse momenta. The radial geometry, however, and thus the volume are
more complicated, similarly to the spherically symmetric volume.

\section{Conclusions}

As discussed in this paper, states and basic operators for symmetric
models can be obtained from full loop quantum gravity in a direct way
and lead to considerable simplifications even in inhomogeneous
models. Hopefully, this will eventually lead to explicit
investigations of important problems in the full theory, such as
general field theory aspects (in particular relating loop to standard
field theory techniques, e.g.\ \cite{AsympSafe,Graviton,QFTonCST}),
issues of the constraint algebra \cite{QSDI,Constr} and
the master constraint \cite{Master}, as well as explicit constructions
of semiclassical states \cite{CohState}. Even though, compared to
homogeneous models, the system is much more complicated with
infinitely many kinematical degrees of freedom, the situation is
simpler than in the full theory.

In addition to the structure of states and basic operators discussed
before, an advantage is that fluxes commute with each other such that
there exists a flux representation. Transforming to such a
representation from the connection representation has been of
significant advantage in homogeneous models, but is not possible in
the full theory with its non-commuting fluxes \cite{NonCommFlux}. Even
so, the Hamiltonian constraint equation will turn into a functional
difference equation with infinitely many independent variables for
which most likely new techniques would have to be developed.

An unexpected complication can arise in inhomogeneous models since
momenta conjugate to the connection may not be identical to triad
components. Thus, even though basic operators are easy to deal with
explicitly, this does not necessarily translate to direct access to
the quantum geometry, most importantly the volume operator. Since the
volume operator also plays an important role in defining the dynamics
\cite{QSDI} and other interesting operators, a complicated volume
operator whose spectrum is not known explicitly would probably render
calculations in the model almost as hard as those in the full
theory. It turns out that the spherically symmetric model still allows
to diagonalize the volume operator explicitly, and to develop an
explicit calculus rather similar to that in homogeneous models
\cite{SphSymmVol}. This fact opens up the possibility of new
conceptual investigations and applications to the physics of black
holes.

\section*{Acknowledgements}

The author is grateful to Hans Kastrup for many discussions and for
initially setting him on the track to spherically symmetric states in
loop quantum gravity. He also thanks Guillermo Mena Marug\'an, Donald
Neville and Madhavan Varadarajan for reigniting his interest in
inhomogeneous models after some time of homogeneous complacency. The
presentation of this work has profited from joint work and discussions
with Abhay Ashtekar and Jurek Lewandowski.

Some of the work on this paper has been done at the ESI workshop
``Gravity in two dimensions,'' September/October 2003. Early stages
were supported in part by NSF grant PHY00-90091 and the Eberly
research funds of Penn State.


\end{document}